\documentclass{article}

\usepackage{amsmath} 
\usepackage{amssymb}
\usepackage{hyperref}
\usepackage{url} 
\usepackage{url}
\usepackage{multirow}
\usepackage{times}
\usepackage{graphicx}
\usepackage{authblk}
\newcommand{\re}{\mathbb R}
{\bf}{\rm}
\newtheorem{thm}{Theorem}[section]

\newcommand{\norm}[1]{\left\lVert#1\right\rVert}
\newcommand{\abs}[1]{\left\lvert#1\right\rvert}
\DeclareMathOperator{\rank}{rank}
\DeclareMathOperator*{\argmin}{arg\,min}

\begin{document}

\title{Classification of geometrical objects  by integrating currents and functional data analysis}

\author{ Sonia Barahona$^{(1)}$, Pablo Centella$^{(1)}$, Ximo Gual-Arnau$^{(2)}$,  Maria Victoria Ib\'a\~nez $^{(3)}$ and Amelia Sim\'o$^{(3)}$ \\
        \small{(1)     Department of Mathematics. Universitat Jaume I.
              Avda. del Riu Sec s/n. 12071-Castell\'on, Spain.}
                           \\
          \small{(2)   Department of Mathematics-INIT. Universitat Jaume I.
              Avda. del Riu Sec s/n. 12071-Castell\'on, Spain.}
                                      \\
         \small{(3)    Department of Mathematics-IMAC. Universitat Jaume I.
              Avda. del Riu Sec s/n. 12071-Castell\'on, Spain.}
    }

\maketitle

\begin{abstract}
This paper focuses on the application of Linear Discriminant Analysis to a set of geometrical objects (bodies) characterized by \emph{currents}.
A \emph{current} is a relevant mathematical object to model geometrical data, like hypersurfaces, through integration of vector fields along them.
As a consequence of the choice of a vector-valued Reproducing Kernel Hilbert Space (RKHS) as a test space to integrate over hypersurfaces, it is possible to consider that hypersurfaces are embedded in this Hilbert space. This embedding enables us to consider classification algorithms of geometrical objects. \\
A method to apply Functional Discriminant Analysis in the obtained vector-valued RKHS is given. This method is based on the eigenfunction decomposition of the kernel.
So, the novelty of this paper is the reformulation of a size and shape classification problem in Functional Data Analysis terms using the theory of currents and vector-valued RKHS.\\
This approach is applied to a 3D database obtained from an anthropometric survey of the Spanish child population with a potential application to online sales of children's wear.

\textbf{keyword} Currents; Statistical Shape and Size Analysis; Reproducing Kernel Hilbert Space; Functional Data Analysis; Discriminant Analysis.

\end{abstract}

\section{Introduction} \label{introduccion}

Supervised classification of geometrical objects (\cite{Ripley07}), i.e. the automated assigning of geometrical objects to pre-defined classes, is a common problem in many scientific fields. This is a difficult task where several challenges have to be addressed.

The first challenge is how to handle  this kind of data from a  mathematical point of view. Several mathematical frameworks have been proposed to deal with geometrical data, and three of them are the most widely used. As a first option, functions can be used to represent closed contours of the objects (curves in 2D and surfaces in 3D) (\cite{Younes98}). Geometrical objects can also be treated as subsets of $\re^n$ (\cite{Serra82,Stoyanetal94}), or finally they can be described as sequences of points that are given by certain geometrical or anatomical properties (landmarks) (\cite{Kendalletal09,dryden2016statistical}).

These approaches are, in general, known as Shape or Size and Shape Analysis and in these settings, the objects are usually embedded into a space which is not a vector space (in  many cases it is a smooth manifold) and on which the geodesic distance as a natural metric is difficult to compute. This makes the definition of statistics particularly difficult; for example, there is no simple, explicit way to compute a mean (\cite{Pennec06, vinue2016k, flores2016intrinsic}).

In our approach, the contour of each geometrical object (curve in $\re^2$, surface in $\re^3$, or hypersurface in $\re^n$), is firstly represented by a mathematical structure named current  (\cite{VaillantGlaunes05,Glaunesetal06}).
Next, each current is associated to an element of a vector valued Reproducing Kernel Hilbert Space (RKHS) by duality  (\cite{Durrleman10, Barahonaetal16}).
This modeling is  weakly sensitive to the sampling of shapes and it does not depend on the choice of parameterizations.

In this paper, we propose a method to apply Discriminant Analysis (\cite{Fisher36}) to this kind of data, i.e. to a sample of curves or surfaces (hypersurfaces in general), which are represented by functions in a vector valued RKHS. For this purpose, Functional Data Analysis (FDA) techniques will be adapted to the case of being working on an RKHS.

The theory of statistics with functional data has become increasingly popular since the end of the 1990s and is now a major field of research in statistics. It is used when the sample space is an infinite-dimensional function space. Although this theory has incorporated many tools from classic parametric or multivariate statistics, the infinite-dimensional nature of the sample space poses particular problems. The books by \cite{SilvermanRamsay05} and \cite{FerratyVieu06} are key references in the FDA literature.

Regarding Functional Discriminant Analysis, \cite{hall01} use functional principal coordinates to reduce the dimension and then apply both a non-parametric kernel and Gaussian-based discriminators to the dimension-reduced data.

Key references on the particular case of Functional Linear Discriminant Analysis (FLDA) are \cite{James01} and \cite{Preda07}.
As an extension of the classical multivariate approach, the aim of FLDA is finding linear combinations such that the between-class variance is maximized with respect to the total variance, but, due to the infinite-dimensional nature of the data, standard LDA can't be used directly. If the functional data have been observed on every point of their domains, one could discretize the domain of the functions to avoid this obstacle. However, doing so usually leads to high-dimensional data that is highly correlated, so this makes estimating the within-class covariance matrix problematic. Usually, the problem can be overcome by using some form of regularization and/or projection of each functional data onto a finite-dimensional space. Then, one can use LDA on the coefficients of the data in this space, as they are a finite-dimensional representation of the infinite-dimensional functional data.

The method that we propose follows the habitual steps of FLDA but for it, several important difficulties have to be overcome. The first problem is given by the fact that, although our data are functional, they are not expressed in the usual way. We propose to solve this problem by using regularization theory (\cite{CuckerSmale01, bickeletal06}).
Secondly, our functions are vector-valued and they are in a vector-valued RKHS.
For this reason we propose to express each vector field  with respect to the orthonormal basis given by the eigenfunction decomposition of the kernel that defines the RKHS. This decomposition can be obtained as a generalization of the similar results to the scalar case (\cite{Quangetal10}).
The coefficients are estimated and the vector associated with each vector field is obtained.
The elements of the basis given by the eigenfunction decomposition of the kernel are orthonormal and ordered following an optimality approximation criterion. These properties allow us to reduce the dimension and then we are able to apply the standard LDA algorithm, as in the multivariate case.

In order to show the applicability of the proposed methodology, we are going to test it on a well known data set of 2D synthetic figures and on a real data set of 3D anthropometric data of Spanish children.
Our implementations have been written in \cite{Matlab15}.

The article is organized as follows:
Section~\ref{Sec:RKHS} concerns the theoretical concepts of currents and Reproducing Kernel Hilbert Spaces.
In Section~\ref{DA} we use the representation of vector fields in the sample with respect to an orthonormal basis in the RKHS to apply the \emph{Discriminant Analysis} algorithm. An experimental study with synthetic figures is conducted in Section~\ref{exp_study2D}.
The application for assigning a size to each child according to his/her body size and shape is detailed in Section~\ref{our_appl}.
Finally, conclusions are discussed in Section~\ref{conclusions}.

\section{Embedding our data in a vector valued Reproducing Kernel Hilbert Space}\label{Sec:RKHS}

As stated in the introduction,  geometrical objects to be classified are embedded in a particular function space.  The motivation for this embedding is the idea of ``currents", which involves characterizing a shape via its ``action" when integrating over vector fields. Seminal papers about currents are \cite{rham60} and \cite{federer60}. This characterization of shapes (mainly curves and surfaces) as currents is studied in detail, for instance, in \cite{Durrlemanetal09}, \cite{Durrleman10} and
\cite{Barahonaetal16}. In this section we revise the main results.

Let $D$ be a compact set in $\re^n$ and let $X$ be the contour of the geometrical object of interest. We assume that $X$ is  a piecewise-defined smooth and oriented hypersurface in $D$ and that $\tau(x)$ is the vector associated with the  $(n-1)$-multivector defined by a basis of the tangent space $T_x X$ (defined almost everywhere).\\

Let $K:D\times D\longrightarrow \re^{n\times n}$ be a matrix valued kernel and  $H_K (D,\re^n)$ its associated vector-valued RKHS with inner product $\langle \,\, , \,\,\rangle_{H_K}$.

We consider the function defined from the integral:
 \begin{equation}
C_X(y)=\int_X K(x,y)  \tau (x) \,dx. \nonumber
\end{equation}

In the particular case where $X$ is a piecewise-defined smooth curve $L$ in $\re^2$, $\tau(x)$ is the tangent vector to $L$ at point $x$ of the curve in $\re^2$. Similarly, if $X$ is a piecewise-defined smooth surface $S$ in $\re^3$, $\tau(x)$ is the normal to the surface $S$ at point $x$.

The ``shape'' of the object will be identified with the function $C_X: D\longrightarrow \re^n$, that is, we deal with hypersurfaces as vector fields in $H_K(D,\re^n)$.

The choice of the kernel determines the vector-valued RKHS, and especially its inner product.
Although it is not known how to choose the ``best'' kernel for a given application (see Appendix B of \cite{Durrleman10}), translation-invariant isotropic matrix kernels, defined from translation-invariant isotropic scalar kernels of the form $k(x,y)= k(\norm{x-y}_{\re^n})$ are often used.

A matrix valued kernel of particular importance is the vector-valued Gaussian kernel:
\begin{eqnarray}\label{funcio22}
K(x,y):= k(x,y) \, {\mathbb I}_{n\times n}=\displaystyle e^{\displaystyle -\frac{\norm{x-y}^2_{\re^n}}{\lambda^2}}{\mathbb I}_{n\times n},
\end{eqnarray}
where ${\mathbb I}_{n\times n}$ is the identity matrix and  $\lambda > 0$ is a scale parameter (bandwidth). This is the matrix valued kernel that we will use in our experiments.\\
Details about the choice of the value of the parameter $\lambda$, and a comparative study between kernels, regarding unsupervised classification, can be found in \cite{Barahonaetal16}.

\subsection{Discrete setting}

In the discrete setting, the vectors $\tau(x)$ are constant over each mesh cell. Then, if $x_j$ is located at the center of mass of mesh cell $j$, and $\tau_j$ is $\tau(x_j)$ scaled by the size of the mesh cell,
\begin{equation}
X \longrightarrow C_X\cong \int_X K(x, \cdot) \tau(x)\,dx\approx \sum_j K(x_j, \cdot) \tau_j, \nonumber
\end{equation}
and if $\varphi_1=\displaystyle\sum_{j=1}^{N_1} K(x^1_j, \cdot)\tau^1_j$ and $\varphi_2=\displaystyle\sum_{j=1}^{N_2} K(x^2_j, \cdot)\tau^2_j$  are two `hypersurfaces', their inner product is

\begin{equation}
 \langle \varphi_1, \varphi_2 \rangle_{H_K}= \displaystyle \sum_{i=1}^{N_1} \sum_{j=1}^{N_2} \tau^1_i \cdot K(x^1_i, x^2_j)\tau^2_j, \nonumber
 \end{equation}
 where $\cdot$ in this expression denotes the inner product in $\re^n$.\\

If $L$ is a planar curve and $\{y_1, y_2, \ldots, y_p\}$ is a discretization of $L$, then $L$ can be represented as a vector field $\sum K(x_j, \cdot)\tau_j$, where  $x_j$ denotes the center of the segment $[y_j, y_{j+1}]$ and $\tau_j$ is the vector $y_{j+1}-y_j$, which is an approximation of the tangent vector.
\vspace{0.2 cm}

If $S$ is a surface in $\re^3$ and  we have a triangulation of $S$, then $S$ can be represented as a vector field $\sum K(x_j, \cdot)\tau_j$, where $x_j$ are the barycenters of the triangles and $\tau_j$ are their area vectors (that is, their unit normal vectors, scaled by their area).

\section{Discriminant Analysis in an RKHS for hypersurface classification} \label{DA}
We have seen in the previous section how to transform geometrical objects into elements of a vector-valued RKHS (a space of vector fields),  therefore, we assume that we have a random sample of size $m$ of vector fields such as:
\begin{equation} \label{span}
\varphi_k(\cdot)= \displaystyle \sum_{i=1}^{N_k} K(x^k_i, \cdot)\tau^k_i\in H_K(D, \re^n), \, \, k=1,\dots, m,
\end{equation}

Because our data are a particular type of vector-valued functions, it seems logical to use the theory of Functional Data Analysis (FDA) and, in particular, Functional Discriminant Analysis for this purpose.

Most applications of FDA are based on applying regularization methods and expressing functional data on an orthonormal basis of functions. This removes the noise, reduces the dimension and allows to use classical multivariate methods.
The use of this type of techniques for our data and for our particular space of functions presents several differences and difficulties.

Firstly, our  data are not expressed in the standard form of functional data. They are defined from the points  $x^k_i$  where the kernel is evaluated (Eq. \ref{span}) and these points are different from one hypersurface (geometrical object) to another.

Secondly, because our space is an RKHS, we must make use of the properties of these spaces to find the most appropriate orthonormal basis in which to project our data. The properties of RKHS are well known and largely used in the scalar case, but our functions are vector-valued, and the vector-valued RKHS are not as well known and many of their properties are still open problems in the functional analysis literature.

These two issues are addressed in the following subsections.



\subsection{Representing the vector fields in the same sample grid of points}\label{Sec:common grid}

Let us consider the vector field $ \varphi_k(\cdot)$ associated with the discretized hypersurface $X_k$ in $D$, given by  Eq.~ (\ref{span}) and being $\{x^k_i\}_{i=1}^{N_k}$ the centers of mass of the respective mesh cells.\\

Let $\{a_i\}_{i=1}^N$ be a sample grid in $D$, the ``Representer Theorem" (\cite{CuckerSmale01}) tell us that for each vector field $\varphi_k$ there exist a unique mapping $\overline{\varphi_k}\colon \re^n \rightarrow \re^n$ such that:
\begin{equation}\label{funcio7}
\overline{\varphi_k}= \argmin_{g \in H_K(D,\re^n)} \frac{1}{N} \sum_{i=1}^N \norm{g(a_i)-b_{k,i}}_{\re^n}^2 +\gamma \norm{g}_{H_K}^2
\end{equation}
where $b_{k,i}:=\varphi_k (a_i)$, $b_{k,i}=( b_{k,i}^1, b_{k,i}^2, \ldots, b_{k,i}^n) \in \re^n$ and $\gamma > 0$ a regularizing parameter.
This way, we can obtain a function that is as smooth as we need (it depends on the choice of $\gamma$) and which fits the data, i.e. $\overline{\varphi_k}(a_i)$ is close to $\varphi_k(a_i)$.

In the experimental study, we choose different values of $\gamma$ and select the appropriate parameter value by checking the performance of the classification.

Moreover, the unique solution $\overline{\varphi_k}$ to Eq.~(\ref{funcio7}) has the expression $\overline{\varphi_k}(\cdot)=\displaystyle \sum_{i=1}^N K(a_i, \cdot)(\beta_{k,i})$, where vectors $\beta_{k,i} \in \re^n$ are obtained by solving the following matrix system:
\begin{eqnarray}
(\gamma \, N \, {\mathbb I}_{N\times N} +K|_a)\beta_k= b_k, \nonumber
\end{eqnarray}
with $K|_a$ the matrix defined as $K|_a(i,j)=k(a_i, a_j) \in \re$, $i,j=1,\ldots, N$, and $\beta_k$, $b_k$ are the following $N \times n$ matrices
\begin{eqnarray*}
\beta_k= \left(\begin{array}{cccc}
\beta_{k,1}^1 & \beta_{k,1}^2 & \ldots & \beta_{k,1}^n\\
\beta_{k,2}^1 & \beta_{k,2}^2 & \ldots & \beta_{k,2}^n\\
\vdots & \vdots & \ldots & \vdots\\
\vdots & \vdots & \ldots & \vdots\\
\beta_{k,N}^1 & \beta_{k,N}^2 & \ldots & \beta_{k,N}^n\\
\end{array}\right); \textup{                }
b_k= \left(\begin{array}{cccc}
b_{k,1}^1 & b_{k,1}^2 & \ldots & b_{k,1}^n \\
b_{k,2}^1 & b_{k,2}^2 & \ldots & b_{k,2}^n\\
\vdots & \vdots & \ldots & \vdots\\
\vdots & \vdots & \ldots & \vdots\\
b_{k,N}^1 & b_{k,N}^2 & \ldots & b_{k,N}^n\\
\end{array}\right).
\end{eqnarray*}

As a result of applying this theorem to all the vector fields of our sample, from now on we will work with a sample of vector fields of the type:
$$\overline{\varphi_k}(\cdot)=\displaystyle \sum_{i=1}^N K(a_i, \cdot)\beta_{k,i} \in H_K(D, \re^n), \, k=1,\dots, m.$$

\subsection{Orthonormal basis in $H_K(D, \re^n)$} \label{Sec:base ortonormal}


In this section we investigate how to obtain an appropriated orthonormal basis of the vector-valued RKHS $H_K(D, \re^n)$ where project our data.  Final projections will be truncated in an optimal way to obtain a finite dimensional approximation, which enables us the use of multivariate discriminant analysis.\\


Let us consider the Hilbert space
$L^2(D, \re^n):= \{f=(f_1,\dots, f_n)\colon D \longrightarrow \re^n \,/\, \norm{f}^2_{L^2(D, \re^n)}= \sum_{i=1}^n \int_D \abs{f_i (x)}^2 \,dx<\infty \}$
and the integral operator, $L_{K,D}\colon L^2(D, \re^n) \longrightarrow  L^2(D, \re^n)$, defined by
\begin{equation}
L_{K,D} f (x) =\int_D K(x,y) (f(y)) \,dy =\left( \int_D k(x,y) f_i(y) \,dy\right)_{i=1}^n =(L_k f_i (x))_{i=1}^n. \nonumber
\end{equation}

\vspace{0.2 cm}

It is easy to check that we have the  conditions to ensure that the integral operator $L_{K,D}$ is self-adjoint, compact, and positive, since each component $L_k$ will be.
\vspace{0.2cm}

Let $\{ \lambda_l\}_{l=1}^{\infty}$ be the eigenvalues of the integral operator $L_k$ and the corresponding eigenfunctions $\{ \phi_l\}_{l=1}^{\infty}$. The spectral theorem implies that $\lambda_1 \geq \lambda_2 \geq \dots \geq 0$ and $\displaystyle \lim_{l \rightarrow \infty} \lambda_l= 0$ (see \cite{Hsing15}).\\

If $\phi_l$ is an eigenfunction of $L_k$ with corresponding eigenvalue $\lambda_l$, then $\psi_l^j= (0,\dots, \phi_l,\dots, 0)$, where $\phi_l$ is placed at position $j$, is an eigenfunction of $L_{K,D}$ corresponding to the same eigenvalue.

\begin{thm} \label{tours}

Let $\overline{\varphi_k}(\cdot)= \displaystyle \sum_{i=1}^N K(\cdot, a_i)(\beta_{k,i})$ be a vector field representing a hypersurface $X_k$ in $\re^n$.
Then, $\overline{\varphi_k}$ can be expressed as
\begin{equation} \label{expresion_base}
\overline{\varphi_k}(\cdot)=\displaystyle \sum_{j=1}^n \sum_{l=1}^{\infty}\mu^j_{l,k} \left(\sqrt{\lambda_l}\psi^j_l (\cdot)\right),
\end{equation}
where $\{\{ \sqrt{\lambda_l} \psi_l^j \}_{l=1}^{\infty}\}_{j=1}^n$ is an orthonormal basis for $H_K(D, \re^n)$.

Furthermore, the first $d= \rank(K|_a)$ coefficients $\mu^j_{l,k}$ can be approximated by
\begin{eqnarray}
\widehat{\mu^j_{l,k}}=\sqrt{\ell_l}(v_l \cdot \beta_k^j)\nonumber
\end{eqnarray}
for $j=1,\ldots, n$, where $v_l \in \re^N$ are the eigenvectors of $K|_a$, $\ell_l$ are the eigenvalues of $K|_a$ and $\beta_k^j= (\beta_{k,1}^j, \beta_{k,2}^j, \ldots, \beta_{k,N}^j)$.
\end{thm}

{\it Proof.}
Consider now a vector field:
\begin{eqnarray} 
\overline{\varphi_k}(\cdot)= \displaystyle \sum_{i=1}^N K(a_i, \cdot)(\beta_{k,i})= \left(\sum_{i=1}^N k(a_i, \cdot)\beta^1_{k,i},\dots, \sum_{i=1}^N k(a_i, \cdot)\beta^n_{k,i}\right). \nonumber
\end{eqnarray}
Then, for $x\in D$, and from  \cite{Quangetal10}, we have
$$\overline{\varphi_k}(x)=\sum_{i=1}^N K(a_i, x)(\beta_{k,i})= \sum_{i=1}^N \sum_{l=1}^{\infty}\lambda_l \phi_l (a_i) \phi_l (x)\beta_{k,i}.$$
Therefore,
$$\begin{aligned}
\overline{\varphi_k}(x)&= \left(\sum_{i=1}^N \sum_{l=1}^{\infty}\lambda_l \phi_l (a_i) \phi_l (x) \beta_{k,i}^1,\dots, \sum_{i=1}^N \sum_{l=1}^{\infty}\lambda_l \phi_l (a_i) \phi_l (x) \beta_{k,i}^n\right)\\
&= \left(\sum_{l=1}^{\infty}\lambda_l\left(\sum_{i=1}^N  \phi_l (a_i) \beta_{k,i}^1\right) \phi_l (x),\dots, \sum_{l=1}^{\infty}\lambda_l\left(\sum_{i=1}^N  \phi_l (a_i) \beta_{k,i}^n\right) \phi_l (x)\right)\\
&= \left(\sum_{l=1}^{\infty}\mu^1_{l,k} \left(\sqrt{\lambda_l}\phi_l (x)\right),\dots, \sum_{l=1}^{\infty}\mu^n_{l,k} \left(\sqrt{\lambda_l}\phi_l (x)\right)\right)
\end{aligned}$$
where $\mu_{l,k}^j= \displaystyle\sqrt{\lambda_l} \sum_{i=1}^N \phi_l (a_i) \beta_{k,i}^j$, $j=1,\dots,n$.
\vspace{0.2 cm}
Again from  \cite{Quangetal10}, $\{\{ \sqrt{\lambda_l} \psi_l^j \}_{l=1}^{\infty}\}_{j=1}^n$ is an orthonormal basis for $H_K(D, \re^n)$; therefore,
for $x\in D$,
$$\overline{\varphi_k}(x)=\sum_{j=1}^n \sum_{l=1}^{\infty}\mu^j_{l,k} \left(\sqrt{\lambda_l}\psi^j_l (x)\right).$$

On the other hand, following the works of \cite{GonzalezMunoz10} and \cite{smale09}, each $\phi_l (a_i)$ can be approximated by $\sqrt{N} v_{l,i}$, where $v_l= \{v_{l,1},\dots, v_{l,N}\}$ is the $l$-th eigenvector of the matrix $K|_a$. In addition, $\lambda_l$ can be estimated by $\ell_l /N$, where $\ell_l$ is the eigenvalue of $K|_a$ corresponding to $v_l$. Therefore
\begin{eqnarray*}
\widehat{\mu_{l,k}^j}= \sqrt{\ell_l}\sum_{i=1}^N v_{l,i} \beta_{k,i}^j= \sqrt{\ell_l} (v_l \cdot \beta_k^j).\quad \square
\end{eqnarray*}

\bigskip

In addition, it is known (\cite[see Theorems~4.4.7 and 4.6.8 in][]{Hsing15}) that for a fixed integer $r>0$ :
\begin{eqnarray*}
\min_{f_1,\dots,f_r\in H_K(D, \re^n)} \int\int_{D \times D}\left( K(y,x)-\sum_{j=1}^n \sum_{l=1}^{\infty} f_l^j(y)f_l^j(x)\right)^2 dydx= \sum_{j=r+1}^\infty \lambda_j^2,
\end{eqnarray*}
where the minimum is achieved by $\sum_{l=1}^r \psi^j_l(y)\psi^j_l(x)$. That is to say, as our functional data are of the form $\sum_{j=1}^n \sum_{l=1}^{\infty}\mu^j_{l,k} \left(\sqrt{\lambda_l}\psi^j_l (x)\right)$,
and the truncated eigenvalue-eigenvector decomposition provides the best approximation to $K$, the truncation of this representation reduces the dimension in an optimal way.

Then, if we truncate the second summation in the Eq. (\ref{expresion_base}) at $d= \rank(K|_a)$, each  hypersurface $X_k$ for $k=1, \ldots, m$, is given by the coefficients $\mu^j_{l,k}$ for $j=1, \ldots, n$ and $l=1, \ldots, d$ (estimated by $\widehat{\mu_{l,k}^j}$), on the orthonormal basis $\{\{ \sqrt{\lambda_l} \psi_l^j \}_{l=1}^{\infty}\}_{j=1}^n$. As a result, it can be represented as the $(n\cdot d)$-dimensional vector
\begin{eqnarray}\label{fapprox}
\mu_k= (\widehat{\mu_{1,k}^1}, \widehat{\mu_{1,k}^2}, \ldots, \widehat{\mu_{1,k}^n},\widehat{\mu_{2,k}^1}, \widehat{\mu_{2,k}^2}, \ldots, \widehat{\mu_{2,k}^n}, \ldots , \widehat{\mu_{d,k}^1}, \widehat{\mu_{d,k}^2}, \ldots\widehat{\mu_{d,k}^n})
\end{eqnarray}
This expression optimally reduces the infinite-dimensional problem to a finite-dimensional one.
Each surface $X_k$ is characterized by the ($n\cdot d$)-dimensional vector through association of currents and now, it is possible to apply the classical multivariate Linear Discriminant Analysis method. In particular, in our applications we will use the multivariate LDA algorithm implemented in \cite{Matlab15}.

%
\section{Experimental 2D Study} \label{exp_study2D}

The public database of synthetic figures ``MPEG7 CE Shape-1 PartB"  (\url{www.imageprocessingplace.com/root_files_V3/image_databases.htm}) contains binary images grouped into categories like cars, faces, watches, horses and birds, with images in the same category showing noticeably different shapes.

In order to illustrate or methodology, three of the categories of this database of synthetic figures are considered: cars, faces and watches. Each class contains $20$ elements, except the watch class, in which two of them are rejected (watch-2 is an atypical element because of its very large size, and watch-8 is considerably tilted and our theoretical framework considers size and shape).
%
In order to have a data set with figures with different sizes, half of the 
figures from each category were enlarged by a scale factor of $1.5$. Then, $10$ of the figures are labelled as "group 1" ("small car"); other $10$ as "group 2" ("large car"); $10$ images are labelled as "group 3" ("small face"), other $10$ as "group 4" ("large face"); $8$ images are labelled as "group 5" ("small watch") and the $8$ remaining images are labelled as "group 6" ("large watch").
After this labelling, each figure of the sample was enlarged by a random coefficient ranging between $1$ and $1.1$, in order to change the ``height" of the figures somewhat.

The $58$ figures were centered, and the contour $\alpha_k$ of each of them defined an oriented smooth curve which was discretized by $100$ points $\{ y_j^k\}_{j=1}^{100}$ ($y_1^k= y_{100}^k$) for $k=1, \ldots, 58$. For each $k\in \{1, \ldots, 58\}$, we defined the centers of the segments
$x_j^k= (y_j^k + y_{j+1}^k)/2$
and the vectors $\tau_j^k= y_{j+1}^k -y_j^k$, $\forall j=1, \ldots, 99$, which define the vector field $\sum_{j=1}^{99} K(x_j^k, \cdot)\tau_j^k$ in $H_K(D, \re^2)$. Gaussian kernels (Eq.~\ref{funcio22}) were used in the definition of the  matrix valued  kernels $K$.

\begin{figure}[htbp]
\begin{center}
\includegraphics[width=7.3cm]{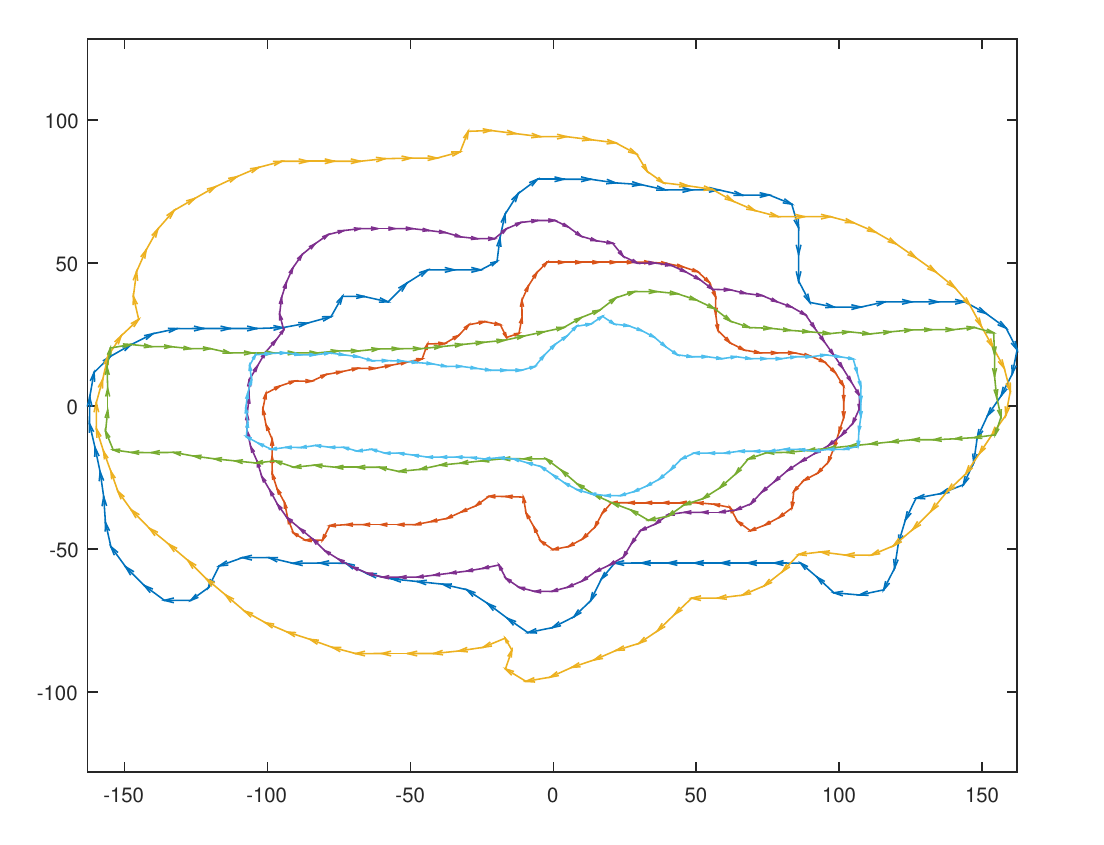}\\
\caption{An object from each class from the synthetic 2D database.  \label{figuras_ejemplo1}}
\end{center}
\end{figure}

Fig.~\ref{figuras_ejemplo1} shows an example of an object from each class. The points $\{ x_j^k \}_{j=1}^{99}$ are plotted in black and the vectors $\{ \tau_j^k \}_{j=1}^{99}$ from each curve are plotted in a different color.

To represent each vector field $\sum_{j=1}^{99} K(x_j^k, \cdot)\tau_j^k$ in relation to a fixed grid using the procedure described in Section~\ref{Sec:common grid}, a grid of $N=3337$ equally spaced points was chosen in $D=[-175, 175] \times [-115, 115]$, with common vertical and horizontal gap $\Delta=5$.

Once we defined the grid, the coefficients of the vector fields that represent each curve in relation to the orthonormal basis of $H_K(D, \re^2)$ were estimated (Section~\ref{DA}).

Different values for parameters $\lambda$ (Eq.~\ref{funcio22}), and $\gamma$ (Eq.~\ref{funcio7}) were chosen. A Functional Discriminant Analysis was conducted and a leave-one-out cross-validation process was carried out to check the performance of the classification obtained, and to select appropriate parameter values.

Table~\ref{tabla_sintetica} shows the results.  As can be seen, results are very satisfactory, as the cross-validation errors are zero for different values of $\lambda$ and $\gamma$. As can be seen in the table, the correct determination of parameter $\lambda$ is more important than the determination of parameter $\gamma$, as $\gamma$ seems less influential in the classification results. The value $67.6$ of the parameter $\lambda$ corresponds to the standard deviation of the points $\{ y_j^k\}_{j=1}^{100}$, $k=1, \ldots, 58$ that define the curves of each sample.

\begin{table}[htbp]
\begin{center}
\begin{tabular}{|c|c|c|}
\hline
$\lambda$ & $\gamma$ &Cross-validation error \\
\hline
    & $1 \times 10^{-4}$ &$30$ ($51.72\%$)\\
 $1$& $3 \times 10^{-4}$ &$30$ ($51.72\%$)\\
    & $4 \times 10^{-4}$ &$30$ ($51.72\%$)\\
\hline
   & $1 \times 10^{-4}$ &$6$ ($10.34\%$)\\
$2$& $3 \times 10^{-4}$ &$6$ ($10.34\%$)\\
   & $4 \times 10^{-4}$ &$6$ ($10.34\%$)\\
\hline
      & $1 \times 10^{-4}$ &$0$ \\
$67.6$& $3 \times 10^{-4}$ &$0$ \\
      & $4 \times 10^{-4}$ &$0$ \\
\hline
     & $1 \times 10^{-4}$ &$0$ \\
$100$& $3 \times 10^{-4}$ &$0$ \\
   & $4 \times 10^{-4}$ &$0$ \\
\hline
\end{tabular}
\caption{Table of results of cross-validation from the 2D database.}
\label{tabla_sintetica}
\end{center}
\end{table}

\section{Application to classify children's body shapes }\label{our_appl}

In 2004, the Biomechanics Institute of Valencia performed an anthropometrical study of the Spanish child population, where a randomly selected sample of Spanish children between the ages of $3$ to $10$ years was scanned using a Vitus Smart 3D body scanner from Human Solutions.


As there is a different size system for each sex, in order to illustrate our procedure the subset of the girls older than 6 was chosen from the whole data set. Children younger than 6 have difficulties in maintaining a standard position during the scanning process, so they were excluded from our data set. The European standard norm UNE-EN 13402-3, defines $4$ different sizes for girls over $6$, (size.6; size.8; size.10 and size.12), corresponding to four consecutive height ranges.
This selection resulted in a sample of size $195$ scanned girls, but in the data set does not provide information about the size that better fits to each of them.

\begin{figure}[htbp]
\begin{center}
\begin{tabular}{c}
\includegraphics[width=10cm]{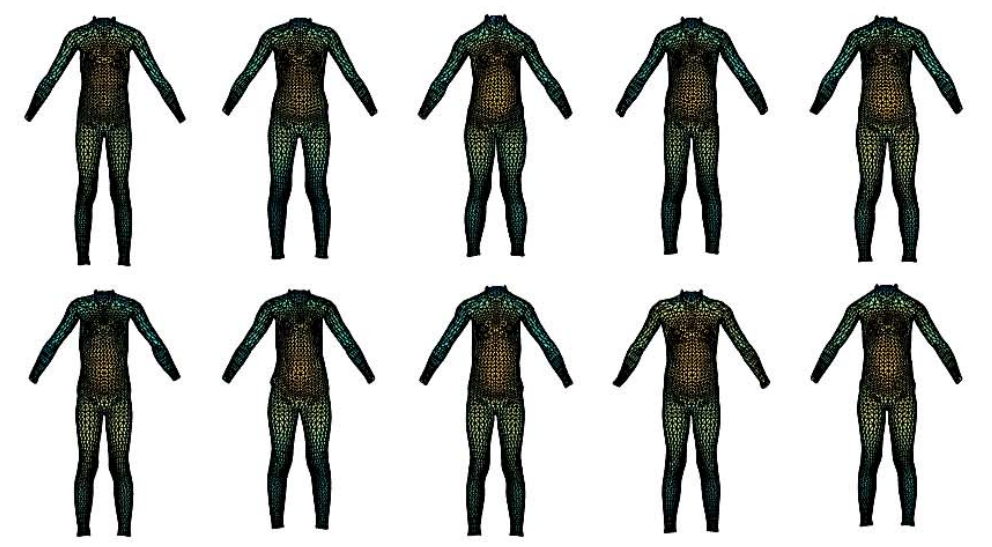}
\end{tabular}
\caption{Some of the girls that have the same size.}
\label{gruponinas}
\end{center}
\end{figure}

In a previous work, \cite{Barahonaetal16} analyzed this data set, and proposed a new sizing system taken into account children's shape and size, without any previous classification by height. This new sizing system defined $5$ different sizes for girls over $6$, instead of the $4$ sizes defined by the standard norm. These sizes will be denoted by $S1$, $\cdots$, $S5$ , and each girl in the data set, was assigned to one of these new sizes. The median (in mm) of the
main anthropometric measurements of the girls of our data set labeled on each size are shown in Table~\ref{tabla_antropometricas_alturas_juntas}.
Fig.~\ref{gruponinas} shows some of the girls labeled as $S1$.

\begin{table}[htbp]
\begin{center}
\begin{tabular}{|c|c|c|c|c|c|}
\hline
Size & Height & Chest & Waist & Hip & Group \\
 & & circumference & circumference & circumference & size \\
\hline
 S1  & $1241$ & $610$ & $540$ & $661$ & $57$ \\ \hline
 S2  & $1259$ & $678$ & $622$ & $754$ & $39$ \\ \hline
 S3  & $1362$ & $660$ & $571.5$ & $723.5$ & $56$ \\ \hline
 S4  & $1361.5$ & $747.5$ & $673$ & $814$ & $34$ \\ \hline
 S5  & $1417$ & $832$ & $767$ & $903$ & $9$ \\ \hline
\end{tabular}
\caption{Median (in mm) of the main antropometric measures of the girls belonging to each size.}
\label{tabla_antropometricas_alturas_juntas}
\end{center}
\end{table}

So, our data set contains a sample of $195$ girls, that have been scanned and assigned to a size (see \cite{Barahonaetal16}), and our aim is to obtain a classification rule that allows us to assign a new girl to her corresponding size, according to this sizing system.

Following the methodology exposed, the body contour from each girl has been represented by an oriented triangulated smooth surface with $4668$ triangles. The center of each oriented triangle with vertices $a_j^k$, $b_j^k$ and $c_j^k$ has been defined as $x_j^k= (a_j^k+b_j^k+c_j^k)/3$, and the normal vectors to the surface are $\tau_j^k= (b_j^k-a_j^k)\times(c_j^k-a_j^k)$, where $j=1,\dots, 4668$ denotes the triangles of the surface and $k=1,\cdots, 195$ denotes each girl. Then, the $k$-th girl's body contour is associated with the vector field $\sum_{j=1}^{4668} K(x_j^k, \cdot)\tau_j^k$ in $H_K(D, \re^3)$, where $D=[-472.73, 487.27]\times[-824.72, 735.28]\times[-156.70,203.30]$  and $\Delta=120$ (in this case, the grid has $N=504$ points)

Once again, a Gaussian kernel (Eq.~\ref{funcio22}) is used in the definition of the  matrix valued  kernel $K$, several suitables values of $\gamma$ are selected, and the value of the parameter $\lambda$ is chosen as the standard deviation of the points $\{x_j^k\}_{j=1}^{4668}$, $k=1, \ldots, 195$ that define the surfaces of each girl.

Using the same procedure as in the 2D database, FLDA was conducted and a leave one out cross-validation procedure was carried out to check the performance of the classification. Table~\ref{tablaninas} shows the results.

\begin{table}[htbp]
\begin{center}
\begin{tabular}{|c|c|c|c|}
\hline
$\lambda$ & $\gamma$ & $\rank(K)$ & CV error \\
\hline
$183.4393$ & $1/930$ & $504$ & $16.92\%$ \\ \hline
\end{tabular}
\caption{Table of results of the cross-validation analysis ($195$ girls).}
\label{tablaninas}
\end{center}
\end{table}

The percentage of misclassifications with this FLDA - cross validation analysis is of $16.92\%$, showing a high agreement percentage ($83.08\%$) in the prediction of the right size of girls' clothing from her body size and shape.
These results, are similar or slightly better than the obtained on the studies found in the literature for similar problems of body shape classification of the adult population (\cite{viktoretal06,devarajanetal04, meunier00}), that use linear anthropometrical dimensions or directly the three-dimensional landmark coordinates as input information; clustering analysis for defining sizing systems  and classical linear discriminant analysis as supervised classification tool.

\section{Discussion \label{conclusions}}

The first aim of this paper has been to adapt and to generalize the linear discriminant analysis methodology for functional data (FLDA) to problems where the data lie on a vector-valued RKHS.

Given a random sample of geometrical objects, each geometrical object (curve or surface), has been represented by a current and each current has been associated to an element (to a function) of a vector-valued RKHS (see Section \ref{Sec:RKHS}).

The main novelty of the work is exposed in Section (\ref{DA}). Most applications of FDA are based on applying regularization methods and expressing functional data on an orthonormal basis of functions. Therefore, as our data (functions in a vector-valued RKHS), are not expressed in the standard form of functional data, the 'Representer Theorem' has been used to define all the functions from the same grid of points (Subsection \ref{Sec:common grid}). Then, it has been necessary to  investigate how to obtain an appropriated orthonormal
basis of the vector-valued RKHS where project our data (Subsection \ref{Sec:base ortonormal}).

To check the performance of this methodology, it has been applied on a supervised classification problem on a data set of synthetic figures  (Section \ref{exp_study2D}). As an example of application, it has been used to get a classifier of child's clothing sizes (Section \ref{our_appl}). Very good results ($0\%$ and  $16.92\%$ cross validation errors respectively) have been obtained in both cases.

\section{Acknowledgments}
This paper has been partially supported by the project $DPI2013-47279-C2-1-R$ from the Spanish Ministry of Economy and Competitiveness with FEDER
funds. 
We would also like to
thank the Biomechanics Institute of Valencia for providing us with
the data set.


\end{document}